\documentclass[letter]{article}

\newcommand{\be}{\begin{equation}}
\newcommand{\ee}{\end{equation}}
\newcommand{\bea}{\begin{eqnarray}}
\newcommand{\eea}{\end{eqnarray}}
\newcommand{\A}{{\mathbf A}}
\newcommand{\X}{{\mathbf X}}

\newcommand{\V}{{\mathbf V}}

\newcommand{\nn}{\nonumber}

\newcommand{\hn}{{\mathbf {\hat n}}}

\newcommand{\half}{\frac{1}{2}}

\begin{document}
\title{Intermediate energy spectrum of five colour QCD at one loop}
\author{M.L. Walker \\
Institute of Quantum Science, Nihon University, \\
Chiyoda, 101-8308, Japan}
\date{}

\maketitle
\abstract{I consider the monopole condensate of five colour QCD. 
The n\"{a}ive lowest energy state is unobtainable at one-loop for five
or more colours due to simple geometry. The consequent
adjustment of the vacuum condensate generates a hierarchy of confinement scales in 
a natural Higgs-free manner. QCD and QED-like forces emerge naturally, acting upon
matter fields that may be interpreted as down quarks, up quarks and electrons.}




\section{Introduction} \label{sec:intro}
It is already known \cite{S77,F80,me07}, that $SU(N)$ QCD 
can lower the energy of its vacuum with
a monopole background field along the Abelian directions, where the Abelian components are
equal in magnitude but orthogonal in real space \cite{F80,me07}. This orthogonality,
while of no special consequence in $SU(3)$ QCD in three space dimensions, does have consequences
when the number of Abelian directions is greater than three.
As noted originally by Flyvbjerg \cite{F80}, $SU(N \ge 5)$ QCD cannot realise its true 
minimum because four orthogonal vectors cannot fit in three
dimensions. I shall call a system kept from reaching its true lowest
energy state by a lack of spatial dimensions \emph{dimensionally frustrated}. 

The research in this chapter seeks to identify the monopole condensate of five-colour QCD, 
or at least a good candidate for it, and examine the consequences. It assumes the dual
superconductor model of confinement \cite{N74,M76,P77,Cho80a,tH81} 
in which chromomagnetic monopole-antimonopole pairs 
play the dual role to Cooper pairs, restricting the electric component of the chromodynamic
field to flux tubes. This model is by no means proven but the case for it is very strong.
I shall handle the monopole degrees of freedom with the Cho-Faddeev-Niemi-Shabanov decomposition
\cite{Cho80a,FN99c,S99b,LZZ00} 
which specifies the internal directions corresponding to the Abelian generators in a
gauge-covariant way, automatically introducing the monopole field in the process. It is 
explained in detail in section \ref{sec:CFNS}. 

At extremely high energies where the effects of confinement are not significant, the dynamics
are simply those of $SU(5)$ QCD in the far ultraviolet. I shall show however that the
dimensionally frustrated condensate is anisotropic and this causes some colours to be confined 
more tightly than others. Even more interesting, white combinations do not necessarily need to
contain all five colours as one would expect. The first three colours, labelled
red, blue and green, form unconfined combinations among themselves, as do the additional two
colours, which I have called ultraviolet and infrared. (Actually the confining effect is not 
quite zero but is much smaller than other effects and is therefore ignored.) This is but one
way in which the $SU(3)$ symmetry naturally breaks off from the rest of the symmetry group.

Examination of the gluon dynamics in section \ref{sec:gluons} 
finds that those corresponding to one particular root vector are
confined more tightly than the rest. At intermediate energies these gluons drop out of the 
dynamics, causing the coupling constants of those that remain to scale differently and again
leading to the separation of $SU(3)$. The emergence of an unconfined Abelian gauge field that
can be identified with the photon is demonstrated in section \ref{sec:photon}. While these
sections are chiefly reviews of work already completed \cite{me07b}, section
\ref{sec:matter} contains new material concerning 
the matter field representations and identifies the neutral,
or white, colour combinations as well as the natural emergence of both up and down quarks
and the electron, all with the correct relative electric and colour charges.

While this is strictly speaking an examination of a one-loop effect in $SU(5)$ QCD, the
prospect of grand unification does arise. The conclusion of this chapter is
that its effective theory does not include weak nuclear decay but that it could well describe
a unification of QCD with QED. The value of such a unification is also discussed along with 
the predictions of this theory in section \ref{sec:predict}, before ending with a summary in 
section \ref{sec:discuss}.

\section{The Cho-Faddeev-Niemi-Shabanov decomposition} \label{sec:CFNS}
My treatment of the monopole condensate rests on the Cho-Faddeev-Niemi-Shabanov
decomposition \cite{Cho80a,FN99c,S99b,LZZ00}. I use the following notation: \newline
The Lie group $SU(N)$ has $N^2-1$ generators $\lambda^{(j)}$, of which $N-1$
are Abelian generators $\Lambda^{(i)}$. 
For simplicity, I specify the 
gauge transformed Abelian directions (Cartan generators)
with 
\be
\hn_i = U^\dagger \Lambda^{(i)} U. 
\ee
In the same way, I replace the standard
raising and lowering operators $E_{\pm\alpha}$ for the root vectors $\mathbf{\alpha}$ 
with the gauge transformed ones
\be
E_{\pm \alpha} \rightarrow U^\dagger E_{\pm \alpha} U,
\ee
where $E_{\pm \alpha}$ refers to the gauge transformed operator 
throughout the rest of this chapter.

Gluon fluctuations in the $\hn_i$ directions are described by $c^{(i)}_\mu$. 
The gauge field of the covariant derivative which leaves
the $\hn_i$ invariant is
\bea
g\mathbf{V}_\mu \times \hn_i = -\partial_\mu \hn_i.
\eea
In general this is 
\bea
\mathbf{V}_\mu = c^{(i)}_\mu \hn_i + \mathbf{B}_\mu ,\; \;
\mathbf{B}_\mu = g^{-1} \partial_\mu \hn_i \times \hn_i,
\eea
where summation is implied over $i$. $\mathbf{B}_\mu$ can be a attributed to
non-Abelian monopoles, as indicated by the $\hn_i$ describing the homotopy group
$\pi_2[SU(N)/U(1)^{\otimes (N-1)}] \approx \pi_1[U(1)^{\otimes (N-1)}]$.
The monopole field strength
\be
\mathbf{H}_{\mu \nu} = \partial_\mu \mathbf{B}_\nu - \partial_\nu \mathbf{B}_\mu
+ g\mathbf{B}_\mu \times \mathbf{B}_\nu,
\ee
has only Abelian components, \textit{ie}. 
\be
H^{(i)}_{\mu\nu}\,\hn_i = \mathbf{H}_{\mu\nu},
\ee
where $H^{(i)}_{\mu\nu}$ has the eigenvalue $H^{(i)}$. Since I am only
concerned with magnetic backgrounds, $H^{(i)}$ is considered the magnitude
of a background magnetic field $\mathbf{H}^{(i)}$. The field strength of the Abelian
components $c_\mu^{(i)}$ also lies in the Abelian directions as expected and is shown
by
\bea
\mathbf{F}_{\mu\nu} = F_{\mu\nu}^{(i)} \hn_i, 
\eea
where
\be
F^{(i)}_{\mu\nu} = \partial_\mu c_\nu^{(i)} - \partial_\nu c_\mu^{(i)}.
\ee
The Lagrangian of the Abelian and monopole components is
\bea \label{eq:Abelian}
-\frac{1}{4} (F_{\mu\nu}^{(i)} \hn_i + \mathbf{H}_{\mu\nu})^2
\eea

The dynamical degrees of freedom (DOF) perpendicular to $\hn_i$ are denoted by
$\X_\mu$, so if $\A_\mu$ is the gluon field then
\bea
\A_\mu &=& \V_\mu + \X_\mu 
= c^{(i)}_\mu \hn_i + \mathbf{B}_\mu + \X_\mu,
\eea
where
\bea
\X_\mu \bot \hn_i , \; \;
\X_\mu = g^{-1}\hn_i \times \mathbf{D}_\mu \hn_i, \; \;
\mathbf{D}_\mu = \partial_\mu + g\A_\mu \times. 
\eea
Because $\X_\mu$ is orthogonal to all Abelian directions it can be expressed as 
a linear combination of the raising and lowering operators $E_{\pm\alpha}$, which
leads to the definition
\begin{eqnarray}
X_\mu^{(\pm \alpha)} \equiv E_{\pm\alpha} \mbox{Tr}[\mathbf{X}_\mu E_{\pm\alpha}],
\end{eqnarray}
so
\begin{equation}
X_\mu^{(-\alpha)} = {X_\mu^{(+ \alpha)}}^\dagger .
\end{equation}

$\mathbf{H}_{\mu \nu}^{(\alpha)}$, defined by
\be
\mathbf{H}_{\mu \nu}^{(\alpha)} = \alpha_j H_{\mu \nu}^{(j)},
\ee
is the monopole field strength tensor felt by $\mathbf{X}_\mu^{(\alpha)}$.
I also define the background magnetic field
\be
\mathbf{H}^{(\alpha)} = \alpha_j \mathbf{H}^{(j)},
\ee
whose magnitude $H^{(\alpha)}$ is $\mathbf{H}_{\mu \nu}^{(\alpha)}$'s non-zero eigenvalue.
Since both $\mathbf{B}_\mu, \X_\mu$ contain off-diagonal degrees of freedom, it is worth 
clarifying that $\X_\mu$ contains the quantum fluctuations taking place on a generally
non-trivial background whose topology is contained in the monopole field $\mathbf{B}_\mu$.

\section{The Vacuum State of five-colour QCD} \label{sec:ansatz}

The one-loop effective energy of five-colour QCD is given by \cite{F80,me07}
\bea
\label{eq:su4ground}
\mathcal{H} = \sum_{\alpha>0} \Vert \mathbf{H}^{(\alpha)} \Vert^2 
\left[\frac{1}{5g^2} + \frac{11}{48\pi^2}
\ln \frac{H^{(\alpha)}}{\mu^2} \right]
\eea
which is minimal when
\bea
H^{(\alpha)} = \mu^2 \exp \left(-\half - \frac{48\pi^2}{55g^2} \right).
\eea
This neglects an alleged imaginary component \cite{NO78} which 
has been called into serious question recently 
\cite{H72,CmeP04,Cme04,CP02,K04,KKP05,me07} with more and more studies finding that
it is only an artifact of the quadratic approximation. Taking this to be the case,
I employ the Savvidy vacuum. This can be criticised for lacking Lorentz covariance
but I argue that it is likely to match the true vacuum at least locally.

Since
\bea \label{eq:rootmagnitudes}
\Vert \mathbf{H}^{(1,0,0,0)} \Vert &=& \Vert \mathbf{H}^{(1)} \Vert ,\nn \\
\Big\Vert \mathbf{H}^{\left(\pm\frac{1}{2},\frac{\sqrt{3}}{2},0,0\right)} \Big\Vert^2 
&=& \frac{1}{4} \Vert \mathbf{H}^{(1)} \Vert^2 + \frac{3}{4} \Vert \mathbf{H}^{(2)} \Vert^2
\pm \frac{\sqrt{3}}{2} \mathbf{H}^{(1)} \cdot \mathbf{H}^{(2)} , \nn \\
\Big\Vert \mathbf{H}^{\left(\pm\frac{1}{2},\frac{1}{\sqrt{12}},\frac{2}{\sqrt{6}},0\right)} \Big\Vert^2 
&=& \frac{1}{4} \Vert \mathbf{H}^{(1)} \Vert^2 + \frac{1}{12} \Vert \mathbf{H}^{(2)} \Vert^2
+ \frac{2}{3} \Vert \mathbf{H}^{(3)} \Vert^2 
\pm \sqrt{\frac{2}{3}} \mathbf{H}^{(1)} \cdot \mathbf{H}^{(2)} \nn \\
&&\pm \frac{1}{2\sqrt{3}} \mathbf{H}^{(1)} \cdot \mathbf{H}^{(3)}
+ \frac{\sqrt{2}}{3} \mathbf{H}^{(2)} \cdot \mathbf{H}^{(3)}, \nn \\
\Big\Vert \mathbf{H}^{\left(0,-\frac{1}{\sqrt{3}},\frac{2}{\sqrt{6}},0\right)} \Big\Vert^2 
&=& \frac{1}{3} \Vert \mathbf{H}^{(2)} \Vert^2 + \frac{2}{3} \Vert \mathbf{H}^{(3)} \Vert^2
- \frac{2\sqrt{2}}{3} \mathbf{H}^{(2)} \cdot \mathbf{H}^{(3)}, \nn \\
\Big\Vert \mathbf{H}^{\left(0,0,-\frac{\sqrt{3}}{\sqrt{8}},\frac{\sqrt{5}}{\sqrt{8}}\right)} \Big\Vert^2 
&=& \frac{3}{8} \Vert \mathbf{H}^{(3)} \Vert^2 + \frac{5}{8} \Vert \mathbf{H}^{(4)} \Vert^2
- \frac{\sqrt{15}}{4} \mathbf{H}^{(3)} \cdot \mathbf{H}^{(4)} , \nn \\
\Big\Vert \mathbf{H}^{\left(0,-\frac{\sqrt{3}}{\sqrt{8}},\frac{1}{\sqrt{24}},\frac{\sqrt{5}}{\sqrt{8}}\right)} \Big\Vert^2 
&=& \frac{3}{8} \Vert \mathbf{H}^{(2)} \Vert^2 + \frac{1}{24} \Vert \mathbf{H}^{(3)} \Vert^2
+ \frac{\sqrt{5}}{\sqrt{8}} \Vert \mathbf{H}^{(4)} \Vert^2 
- \sqrt{\frac{1}{16}} \mathbf{H}^{(2)} \cdot \mathbf{H}^{(3)} \nn \\
&&- \frac{\sqrt{15}}{4} \mathbf{H}^{(2)} \cdot \mathbf{H}^{(4)}
+ \frac{\sqrt{5}}{\sqrt{48}} \mathbf{H}^{(3)} \cdot \mathbf{H}^{(4)}, \nn \\
\Big\Vert \mathbf{H}^{\left(\pm\frac{1}{2},\frac{1}{\sqrt{12}},\frac{1}{\sqrt{24}},
\frac{\sqrt{5}}{\sqrt{8}}\right)} \Big\Vert^2 
&=& \frac{1}{4} \Vert \mathbf{H}^{(1)} \Vert^2 + \frac{1}{12} \Vert \mathbf{H}^{(2)} \Vert^2
+ \frac{1}{24} \Vert \mathbf{H}^{(3)} \Vert^2 + \frac{\sqrt{5}}{\sqrt{8}} \Vert \mathbf{H}^{(4)} \Vert^2 \nn \\
&&\pm \sqrt{\frac{2}{3}} \mathbf{H}^{(1)} \cdot \mathbf{H}^{(2)} 
\pm \frac{1}{\sqrt{24}} \mathbf{H}^{(1)} \cdot \mathbf{H}^{(3)}
+ \frac{1}{\sqrt{16}} \mathbf{H}^{(2)} \cdot \mathbf{H}^{(3)} \nn \\
&&\pm \frac{\sqrt{5}}{\sqrt{8}} \mathbf{H}^{(1)} \cdot \mathbf{H}^{(4)}
+ \frac{\sqrt{5}}{\sqrt{24}} \mathbf{H}^{(2)} \cdot \mathbf{H}^{(4)}
+ \frac{\sqrt{5}}{\sqrt{48}} \mathbf{H}^{(3)} \cdot \mathbf{H}^{(4)},
\eea
it follows that 
\be
\Vert \mathbf{H}^{(i)}\Vert = \Vert \mathbf{H}^{(j)}\Vert ,\;\;
\mathbf{H}^{(i)} \bot \mathbf{H}^{(j)} ,\;\; i\ne j,
\ee
which means that
the chromomagnetic field components must be equal in magnitude but
mutually orthogonal in the lowest energy state. However three dimensional space can
only accomodate three mutually orthogonal vectors. Since the number of Cartan components,
\textit{ie.} components corresponding to Abelian generators,
is always $N-1$ in $SU(N)$ it follows that QCD with more than four colours cannot 
achieve such an arrangement.

One could substitute the Cartan basis $\mathbf{H}^{(i)}$ but this leads
to intractable equations that cannot be solved analytically. It is reasonable to expect that
the lowest attainable energy state is only slightly different from (\ref{eq:su4ground}) and
that this difference is due to the failure of mutual orthogonality. I therefore propose the
ansatz that all Cartan components are equal in magnitude to what they would be in the absence
of dimensional frustration, and that their relative orientations in real space are chosen so as
to minimise the energy. In practice this means that three of the four are mutually orthogonal
and the remaining one is a linear combination of those three. This remainder will increase the 
effective energy through its scalar products with the mutually orthogonal vectors but not all
scalar products contribute equally. This follows from the form of the root vectors in 
eq.~(\ref{eq:rootmagnitudes}). This means that the orientation of the remaining real space
vector in relation to the mutually orthogonal ones impacts the effective energy. 

A little 
thought
reveals that the lowest energy state should have only one scalar product contribute to it. The
problem of finding the lowest available energy state therefore reduces to finding the scalar product 
that contributes to it the least. The six candidates are
\bea \label{eq:candidates}
\mathbf{H}^{(1)} \cdot \mathbf{H}^{(2)} ,\mathbf{H}^{(1)} \cdot \mathbf{H}^{(3)} ,\mathbf{H}^{(1)} \cdot \mathbf{H}^{(4)} ,\nn \\
\mathbf{H}^{(2)} \cdot \mathbf{H}^{(3)} ,\mathbf{H}^{(2)} \cdot \mathbf{H}^{(4)} ,\mathbf{H}^{(3)} \cdot \mathbf{H}^{(4)} .
\eea
\begin{table}
	\centering
	\caption{Candidate parallel components for vacuum condensate. The column on the left 
is for parallel vectors, the column on the right is for antiparallel vectors.
$\Delta \mathcal{H}$ should be multiplied by $H^2 \frac{11}{96\pi^2}$.}
		\begin{tabular}{cc|cc} \hline \hline
			$\mathbf{H}^{(i)} = + \mathbf{H}^{(j)}$ & $\Delta \mathcal{H}$ 
			& $\mathbf{H}^{(i)} = - \mathbf{H}^{(j)}$ & $\Delta \mathcal{H}$ \\ \hline
			$\mathbf{H}^{(1)} = + \mathbf{H}^{(2)}$ & $1.06381$ &
			$\mathbf{H}^{(1)} = - \mathbf{H}^{(2)}$ & $1.06381$ \\
			$\mathbf{H}^{(1)} = + \mathbf{H}^{(3)}$ & $0.857072$ &
			$\mathbf{H}^{(1)} = - \mathbf{H}^{(3)}$ & $0.857072$ \\
			$\mathbf{H}^{(1)} = + \mathbf{H}^{(4)}$ & $0.715651$ &
			$\mathbf{H}^{(1)} = - \mathbf{H}^{(4)}$ & $0.715651$ \\
			$\mathbf{H}^{(2)} = + \mathbf{H}^{(3)}$ & $1.01655$ &
			$\mathbf{H}^{(2)} = - \mathbf{H}^{(3)}$ & $0.656584$ \\
			$\mathbf{H}^{(2)} = + \mathbf{H}^{(4)}$ & $0.882589$ &
			$\mathbf{H}^{(2)} = - \mathbf{H}^{(4)}$ & $0.577976$ \\
			$\mathbf{H}^{(3)} = + \mathbf{H}^{(4)}$ & $1.00042$ &
			$\mathbf{H}^{(3)} = - \mathbf{H}^{(4)}$ & $0.540983$ \\ \hline
		\end{tabular}
	\label{tab:Candidates}
\end{table}
As can be seen from table \ref{tab:Candidates},
$\mathbf{H}^{(3)} = -\mathbf{H}^{(4)}$ (antiparallel) 
yields the lowest effective energy when all other scalar products are zero. 

Substituting this result into (\ref{eq:rootmagnitudes}) finds that all $\mathbf{H}^{(\alpha)}$ 
have the same magnitude except for those that couple to $\mathbf{H}^{(4)}$, namely
$\mathbf{H}^{\left(?,?,?,\sqrt{\frac{5}{8}}\right)}$, 
where ? indicates that there are several possible values.
The other background field strengths are 
\be
\Vert \mathbf{H}^{(\alpha)} \Vert^2 = H^2,
\ee
while the strongest is
\be
\Vert \mathbf{H}^{\left(0,0,-\sqrt{\frac{3}{8}},\sqrt{\frac{5}{8}}\right)} \Vert^2 
= H^2 \left(1+\frac{\sqrt{15}}{4}\right),
\ee
and the weakest are
\be
\Vert \mathbf{H}^{\left(?,?,\frac{1}{\sqrt{24}},\sqrt{\frac{5}{8}}\right)} \Vert^2 
= H^2 \left(1-\sqrt{\frac{5}{48}} \right).
\ee
Remember that the negative signs are affected by $\mathbf{H}^{(3)},\mathbf{H}^{(4)}$ being 
antiparallel.

Assuming the dual superconductor model of confinement 
\cite{N74,M76,P77,Cho80a,tH81},
it follows that different valence gluons and even different quarks (in the fundamental representation)
will be confined with different strengths and therefore at different length scales. Those that feel the 
background $H^{\left(0,0,-\frac{\sqrt{3}}{\sqrt{8}},\frac{\sqrt{5}}{\sqrt{8}}\,\right)}$ will 
be confined the most strongly, those that feel the backgrounds of the form
$H^{\left(?,?,\frac{1}{\sqrt{24}},\frac{\sqrt{5}}{\sqrt{8}}\,\right)}$ will be confined
least strongly. The remainder will be confined with intermediate strength.

At highest energy then, we have the full dynamics of $SU(5)$ QCD. Moving down to some intermediate energy
however, finds that the dynamics associated with the root vector
$\left(0,0,-\frac{\sqrt{3}}{\sqrt{8}},\frac{\sqrt{5}}{\sqrt{8}}\right)$
are confined out of the dynamics. The remaining
gluons interact among themselves. Moving to lower energy scales I find that those
dynamics are all removed in their turn except for those corresponding to the root vectors
$\left(?,?,\frac{1}{\sqrt{24}},
\frac{\sqrt{5}}{\sqrt{8}}\right)$, almost leaving an $SU(2)$ gauge field interaction.
I say 'almost' because I shall later demonstrate that the form of the monopole condensate
is sufficiently different from the $SU(2)$ condensate to alter the dynamics, producing three
confined $U(1)$ gauge fields, two of which are contained within $SU(3)$, a further unconfined
$U(1)$ gauge field that may be identified with the photon, and three copies of 
the valence gluons of $SU(2)$. At lowest energies only the unconfined gauge field remains.
In this way a hierarchy of confinement scales and effective dynamics emerges
naturally, without the introduction of any \textit{ad.~hoc.} mechanisms like the
Higgs field.

\section{Intermediate Energy Dynamics} \label{sec:gluons}
In constructing the hierarchical picture above, I began with $SU(5)$ and finished
with $U(1)$ but had no apparent gauge group governing the dynamics
at the intermediate energy scale. The dynamics of this energy scale will prove to be 
quite interesting. 

To facilitate the discussion I introduce a notation inspired by the Dynkin diagram
of $SU(5)$. The root vectors implicitly specified in eq.~(\ref{eq:rootmagnitudes}) are
all linear combinations of a few basis vectors, which according to Lie algebra representation
theory can be chosen for convenience. I take the basis vectors 
\bea
(1,0,0,0),\left(-\half,\frac{\sqrt{3}}{2},0,0 \right),
\left(0,-\frac{1}{\sqrt{3}},\sqrt{\frac{2}{3}},0 \right),
\left(0,0,-\sqrt{\frac{3}{8}},\sqrt{\frac{5}{8}}\right),
\eea
which I shall each represent by 
\be
\mbox{OXXX},\mbox{XOXX},\mbox{XXOX},\mbox{XXXO},
\ee
respectively. The remaining root vectors are sums of these basis vectors. In this notation
their representation contains an 'O' if the corresponding basis vector is included and
'X' if it is not. For example the root vector
\be
\left(\half,\frac{\sqrt{3}}{2},0,0 \right) 
= (1,0,0,0) + \left(-\half,\frac{\sqrt{3}}{2},0,0\right),
\ee
is represented by
\be
\mbox{OOXX}=\mbox{OXXX}+\mbox{XOXX}.
\ee
When convenient, a '?' is used to indicate that either 'O' or 'X' might be substituted.

In addition to its brevity, this notation has the nice feature of making obvious which
root vectors can be combined to form other root vectors because there are
no root vectors with an 'X' with 'O's on either side. There is no OXXO for example.

The confinement of $X_\mu^{\left(0,0,-\frac{\sqrt{3}}{\sqrt{8}},
\frac{\sqrt{5}}{\sqrt{8}}\right)}$, the valence gluon corresponding to XXXO, 
out of the dynamics 
directly affects only those remaining valence gluons that couple to it, those of root vectors of
the form ??O?. The remaining gluons, corresponding to 
the root vectors OOXX, XOXX and OXXX (collectively given by ??XX), 
may still undergo the full set of
interactions available to them at higher energies. It is easy to see that these are the
root vectors that comprise the group $SU(3)$, to which the other valence gluons couple
forming two six dimensional representations. 
Subsequent discussion shall
extend the X,O,? notation to include the valence gluons corresponding to a root vector.
Whether it is the gluon or the root vector that is meant will be clear from context.

Consider the beta function, or to be less imprecise, the 
scaling of the various gluon couplings. I shall now demonstrate that the loss to 
confinement of the root vector XXXO causes unequal corrections to the running of the
couplings for different gluons. 
Since this is only an introductory paper the following analysis is
only performed to one-loop.

The gluons ??XX, corresponding to the above-mentioned $SU(3)$, retain their original 
set of interactions. Performing the standard perturbative calculation \cite{F80}
therefore yields the standard result for $SU(5)$ QCD. The remaining gluons do not.
The absence of the maximally confined XXXO restricts their three-point vertices to those of
$SU(4)$, since all root vectors are now of the form ???X.
The same is not true of the four-point interactions, but the exceptions do
not contribute to the scaling of the coupling constant at one-loop \cite{Fbook87}. 
We have then that the
$SU(3)$ subgroup's coupling scales differently from the rest of the unconfined gluons when the
maximally confined valence gluons XXXO drop out.

The beta function is proportional to the number of colours in pure QCD at one-loop,
so as the length scale increases, the coupling among gluons within the $SU(3)$ subgroup
initially grows faster than the couplings involving the other gluons. As noted above,
the $SU(3)$ couplings will initially scale as in the five-colour theory, 
while the remainder scale as though there were only four colours.
This specific behaviour must soon change due to both non-perturbative contributions
and because the non-$SU(3)$ gluons have a weaker coupling. A detailed understanding
requires a nonperturbative analysis well beyond the scope of this chapter. 
Indeed, the application of one-loop perturbation 
theory at anything other than the far ultraviolet is questionable in itself. The point
remains that the $SU(3)$ subgroup ??XX separates from the remaining 
gluons by its stronger coupling strength.

The symmetry reduction that takes place in this model is 
suggestive of boson mass generation but there appears to be no obvious
specific mechanism. Kondo \textit{et.~al.} have argued for the spontaneous generation
of mass through various non-trivial mechanisms \cite{K04,K06,me07}. This is consistent
with the well-studied correlation between confinement and chiral symmetry breaking
(see \cite{HF04,KLP01,YS82,PW84} and references therein).

\section{The emergence of QED} \label{sec:photon}
Neglecting off-diagonal gluons, 
the equality $\mathbf{H}^{(3)}=-\mathbf{H}^{(4)}$ allows the change in variables
\bea \label{eq:AandZ} 
c_\mu^{(3)}\hn_3 \rightarrow \half (c_\mu^{(3)}\hn_3 + c_\mu^{(4)}\hn_4) 
+ \half (c_\mu^{(3)}\hn_3 - c_\mu^{(4)}\hn_4) 
= \frac{1}{\sqrt{2}}(A_\mu + E_\mu) ,\nn \\
c_\mu^{(4)}\hn_4 \rightarrow \half (c_\mu^{(3)}\hn_3 + c_\mu^{(4)}\hn_4) 
- \half (c_\mu^{(3)}\hn_3 - c_\mu^{(4)}\hn_4)
= \frac{1}{\sqrt{2}}(A_\mu - E_\mu ).
\eea
Substituting eqs~(\ref{eq:AandZ}) into the Abelian dynamics (\ref{eq:Abelian})
finds that the antisymmetric combination $E_\mu$
couples to the background 
\begin{displaymath}
H (\hn_3 - \hn_4),
\end{displaymath}
but the symmetric combination $A_\mu$
does not. Again by the dual superconductor model the former is confined (along with
$c_\mu^{(1)}\hn_1,c_\mu^{(2)}\hn_2$) while the latter
is not. Since the electromagnetic field is
long range it is natural to interpret $A_\mu$ as the photon.

The rotation from $c_\mu^{(3)}\hn_3,c_\mu^{(4)}\hn_4$ to 
$E_\mu,A_\mu$ in interactions with
valence gluons is only meaningful if the gluon in question couples either to both of
$c_\mu^{(3)}\hn_3$ and $c_\mu^{(4)}\hn_4$ or to neither of them. 
Otherwise the combination of $E_\mu$ and 
$A_\mu$ is ill-defined because it is not unique, \textit{ie.} if the valence gluon  
couples to $c_\mu ^{(3)}\hn_3$ but not to $c_\mu^{(4)} \hn_4$ then arbitrary multiples of
$c_\mu^{(4)} \hn_4$ may be added to the interaction term, yielding
arbitrary mixtures of $E_\mu$ and $A_\mu$. The 
gluons for which this occurs have root vectors of the form ??OX and their electric charge
is ill-defined. This is of little consequence in practice because we shall see in section 
\ref{sec:matter} that such gluons have no sources at intermediate energies.

\section{Matter field representations} \label{sec:matter}
I have shown how the coupling of the gluons to the monopole background determines
their confinement strength and subsequent phenomenology. I now consider the matter
fields and focus in particular on the fundamental representation of $SU(5)$. The 
confinement of the fundamental representation is determined by the maximal
stability group \cite{KT00a,KT00b}, which for $SU(5)$ is $U(4) \approx SU(4) \otimes U(1)$,
where for any given element $( \cdots \, \psi \, \cdots )^T$,
the $SU(4)$ acts only on the remaining orthogonal elements while the $U(1)$ causes it
inconsequential phase changes. This latter $U(1)$ describes the monopole condensate 
contributing to the confinement and is given by the corresponding weight of the 
fundamental representation. As a concrete example, consider the fundamental element
$( 0 \, 0 \, 0 \, 0 \, \psi)^T$.
The $SU(4)$ of its maximal stability group are the matrices
\bea
\left( \begin{array}{cc}
T_i & \begin{array}{c} 0 \\ 0 \\ 0 \\ 0 \end{array} \\
0\; 0\; 0\; 0 & 1
\end{array} \right),
\eea
where $T_i$ are the standard $SU(4)$ matrices $T_1\ldots T_{15}$, while the $U(1)$
is generated by $T_{24} = \frac{1}{\sqrt{20}}\, diag (1\; 1\; 1 \; 1 \; -4 )$.
Therefore the only component of the chromomonopole condensate that contributes to the
confinement of $( 0 \, 0 \, 0 \, 0 \, \psi)^T$ is that generated by $T_{24}$. This is
what would be expected based on the weights of the fundamental representation, and 
indeed the main result of \cite{KT00b} is that the weight of the representation determines
which components of the monopole condensate contribute to a given particle's confinement.

The weights of the fundamental representation of $SU(5)$ are 
\bea \label{eq:weights}
( 1 \, 0 \, 0 \, 0 \, 0)^T & : & 
\left(\half , \frac{1}{\sqrt{12}} , \frac{1}{\sqrt{24}} , \frac{1}{\sqrt{40}}\right) \nn \\
( 0 \, 1 \, 0 \, 0 \, 0)^T & : & 
\left(-\half , \frac{1}{\sqrt{12}} , \frac{1}{\sqrt{24}} , \frac{1}{\sqrt{40}}\right) \nn \\
( 0 \, 0 \, 1 \, 0 \, 0)^T & : & 
\left( 0 , -\frac{1}{\sqrt{3}} , \frac{1}{\sqrt{24}} , \frac{1}{\sqrt{40}} \right) \nn \\
( 0 \, 0 \, 0 \, 1 \, 0)^T & : & 
\left( 0 , 0 , -\sqrt{\frac{3}{8}} , \frac{1}{\sqrt{40}} \right) \nn \\
( 0 \, 0 \, 0 \, 0 \, 1)^T & : & 
\left( 0 , 0 , 0, -\sqrt{\frac{2}{5}} \right)
\eea
If all Abelian components of the chromomonopole condensate were of equal
magnitude and mutually orthogonal in real space so that cross-terms could be neglected
then they would all be confined at equal length scales and nothing remarkable would
happen. However, we already know that such is not the case.

First consider the first three lines in equation (\ref{eq:weights}). They all have identical 
(small) dependence on $\mathbf{H}^{(3)},\mathbf{H}^{(4)}$, and the same dependencies on
$\mathbf{H}^{(1)},\mathbf{H}^{(2)}$ as in $SU(3)$ QCD. Consider now that the last two lines
show no dependence on $\mathbf{H}^{(1)},\mathbf{H}^{(2)}$, and it is only natural to
equate the first three elements with the quark colours of the standard model. 
This is supported by noting that the two additional
weight entries would provide very little additional confinement 
because the cross terms between $\mathbf{H}^{(3)},\mathbf{H}^{(4)}$ have negative sign
due to their antiparallelism. In fact the total contribution squared of 
$\mathbf{H}^{(3)},\mathbf{H}^{(4)}$ is
\be
\left(\frac{1}{\sqrt{24}}\mathbf{H}^{(3)}+\frac{1}{\sqrt{40}}\mathbf{H}^{(4)}\right)^2 
= H^2 \frac{1}{60} (4 - \sqrt{15}).
\ee
As can be seen from the bracket on the right-hand-side, 
the cross terms almost cancel this contribution
entirely. According to the n\"{a}ive interpretation of the dual superconductor model
employed by this paper this corresponds to extremely weak confinement. Since it is
inconsequential compared to the QCD confinement and might very well scale to zero at
lower energies anyway (although I have not shown this!) I shall assume
that this corresponds to an unconfined state.

This all ties in rather nicely with the result of section \ref{sec:gluons} in which the 
corresponding $SU(3)$ dynamics separate from the remaining dynamics through stronger 
scaling of the coupling constant.

The remaining weights have non-zero elements only in the third and fourth position. 
The final weight corresponds to a colour charge which I shall refer to as infrared ($i\!r$),
whose confinement is slightly stronger than that 
of the QCD colours discussed above, while the penultimate one corresponds to the colour 
charge ultraviolet ($u\!v$), 
whose confinement is nearly twice as strong as that of the QCD colours. (I shall refer to 
both of these charges as the invisible colours.) This occurs because there are
positively contributing cross terms between $\mathbf{H}^{(3)},\mathbf{H}^{(4)}$ (remember their 
antiparallelism). It follows that ultraviolet must be combined into some
neutral combination with infrared at a very small length scale. Only combined with 
infrared does ultraviolet form an unconfined physical state. 

Note that the third and fourth entries of the sum of the ultraviolet and infrared weights
are exactly negative three times those of the QCD quark weights.
We have already seen that such a combination effectively feels no confining effect
from the background condensate. Remembering that the third and fourth Abelian directions
provide the unconfined photon $A_\mu$ of section \ref{sec:photon}
gives the electric charge ratio between QCD quark states and white states. In
conventional QCD the white states comprise both white combinations of QCD quark colours
(hadrons) and truly colourless particles (leptons), and it is simply a fortunate coincidence
that both have integer multiples of the electron charge. In this model, states carrying both
of the invisible colours but no QCD colours are white and electrically
charged, as are white combinations of the QCD colours. From the above discussion of the weights
it follows that both these cases also have the same electric charge, up to a negative sign.
It will therefore be natural to interpret the state with both invisible colours but no QCD 
colours as the electron/positron.

The reader may recall that the third and fourth Abelian directions also
provide the confined photon $E_\mu$. This may confuse some. While it may provide some
additional interactions at close range, $E_\mu$ is \emph{confined}, not \emph{confining},
so the electric charges are still freeto separate to very large distances.

Based on the proceeding discussion the fundamental representation can be shown as
\be 
[1] = (r \, b \, g \, u\!v \, i\!r )^T,
\ee
where the electric charge of the QCD quarks is implicit in the colour. The red, blue and
green quark colours are exchanged by the corresponding $SU(3)$ but the invisible colours
are not exchanged except at extremely high energies due to the extra-strong confinement
of the ultraviolet charge. At highest energies the dynamics are
those of $SU(5)$ in the extreme weak-coupling limit. Electric charge has no meaning at
such energies. 

At intermediate energies at which the gluons XXXO have been confined out
of the dynamics not only is there no available gluon to exchange the invisible colours,
but the ultraviolet colour itself is confined just as QCD colours are confined at larger
distances. This not only removes ultraviolet from the effective dynamics but the
infrared as well, because the all-white combinations involving the infrared, apart from a 
meson-like bound state, require the ultraviolet. This confinement of infrared leaves no
source for gluons of the form ??OX which, as discussed in section \ref{sec:gluons}, 
have an ill-defined electric charge. They may still occur in gluon-antigluon pairs but this
is no threat to electric charge. 

The intermediate dynamics consist primarily of
conventional QCD and QED, but there is no obvious way to include the $SU(2)$ of weak nuclear
decay or to turn a QCD quark into a lepton.

I now turn to the asymmetric representation
\be
[2] = [1] \otimes_{AS} [1] = \left[ \begin{array}{ccccc}
0 & r/b & r/g & r/u\!v & r/i\!r \\
-r/b & 0 & b/g & b/u\!v & b/i\!r \\
-r/g & -b/g & 0 & g/u\!v & g/i\!r \\
-r/u\!v & -b/u\!v & -g/u\!v & 0 & u\!v/i\!r \\
-r/i\!r & -b/i\!r & -g/i\!r & -u\!v/i\!r & 0 
\end{array} \right],
\ee
where $[1]$ is the fundamental representation and $\otimes_{AS}$ indicates an antisymmetric
cross-product. 
The top-left-hand corner has the same interpretation as in conventional GUT theories 
\cite{G99book}.
Red/blue is effectively antigreen etc, and the $U(1)$ (in this case electric) charge
is double that of the QCD quarks in the fundamental representation. In other words the
$3 \times 3$ block matrix in the top-left-hand corner can be associated with the anti-up
quark when the fundamental representation contains the down quark. The remaining entries
of $[2]$ contain either an ultraviolet or an infrared colour charge, which confines them
out of intermediate energy level dynamics. The one exception 
contains both invisible colour charges and is therefor a colourless state with electric
charge negative three times that of the down quark, \textit{ie.} 
a positron as discussed above. The effective dynamics of 
this representation, and its complex conjugate $[3]$, are dominated by the colour and electric
interactions of the up quark and electric interactions of the electron/positron.

\section{Predictions and prospects for grand unification} \label{sec:predict}
This approach yields no weak interaction dynamics but there is a gluon-mediated exchange between
the up quarks and the electron. This is reminiscent of proton decay in conventional
GUTS, in which a down quark becomes a positron and an up quark becomes an anti-up quark
which forms a meson with the remaining up quark. In this case however, an up quark becomes 
an electron and the mediating gluon carries a charge of $+\frac{5}{3}$. This cannot be absorbed 
by anything within the proton so proton decay is forbidden. It could however be
absorbed by an anti-up quark so that an extremely high energy collision between a proton and
an anti-proton yielding an electron-positron pair and a neutral pair of pions, by which
I mean either two $\pi^0$s or one $\pi^+$ and one $\pi^-$. Unfortunately such a sequence
can also occur through ordinary standard model interactions so this is not much of a prediction.

It was natural to hope that dimensional frustration might yield a Higgsless GUT
but it makes a good start, unifying $SU(3)_c$ with $U(1)_{EM}$, the forces acting on
the right-handed matter fields, which do not feel the weak nuclear force. While
I cannot yet claim to have done so, a physically realistic unification of $SU(3)_c$ with
$U(1)_{EM}$ would be a unification of the forces affecting right-handed matter fields.

Assuming that nature does employ this mechanism to unify the strong 
nuclear and electromagnetic interactions of right-handed particles, what new
phenomena could we expect to see? Obviously there would be new hadrons
containing the invisible quarks. These can be divided into two types. There are mesons,
we could call them invisible mesons, composed of invisible quark-antiquark pairs. 
Remember that such pairs can be taken not just from the fundamental representation which
contains pure invisible states, but also from the $[2]$ representation which contains
states carrying both a QCD colour and an invisible colour. From QCD colour symmetry 
such states would be mixed, so the distinguishable invisible mesons are
\be \label{eq:invisiblemeson}
u\!v-\overline{u\!v},\; i\!r-\overline{i\!r}, \;u\!v/\{rbg\}-\overline{u\!v/\{rbg\}},\; i\!r/\{rbg\}-\overline{i\!r/\{rbg\}}.
\ee
It is, of course, possible that these states also mix, either with each other or with 
standard model mesons. This requires further study and is beyond the scope of this paper.

There is one more invisible meson which is listed separatedly because it is
already known. Recall that the positron is the quark in the $[2]$ representation
with the colour charge $u\!v/i\!r$. The invisible meson associated with the positron
is obviously positronium, so
\be \label{eq:positronium}
e^+-e^- \Longleftrightarrow  u\!v/i\!r-\overline{u\!v/i\!r}.
\ee

Another obvious combination is the quark pair made up of each invisible colour from
the fundamental representation, $u\!v-i\!r$, and of course its antimatter partner.
Next there are particles made up of those quarks in $[2]$ that carry both a
QCD colour and an invisible colour, \textit{ie.} $\{rbg\}/u\!v$ and $\{rbg\}/i\!r$. An
unconfined combination needs each QCD colour in equal numbers and each of the invisible 
colours in equal numbers, so at least six quarks are needed.

The quark combinations in the last paragraph are neither mesonic nor baryonic and may
be candidates for dark matter. There may also be fancier combinations involving more 
quarks or even gluons similar to the exotic states discussed in relation to conventional QCD.

Conventional GUTs predict proton decay and monopole production. Dimensional frustration
predicts neither of these. It does predict an anomolous scattering however. If an electron 
is fired into a proton at sufficiently high energies it may turn into an up quark and
emit a mediating boson that is absorbed by an up quark and turns it into an electron. The
experimenter sends in an electron and sees an electron emerge so this is just a scattering
experiment, but it is additional scattering to the electromagnetic interaction already
observed in deep inelastic scattering experiments. 
The strength of the scattering for a given electron energy has not yet been calculated and
requires the mediating gluon mass which is currently unknown, although it is natural to expect
that it is very massive so that extremely high energy scattering
experiments would be needed. It is worth noting however that while reaching ever higher 
energies is becoming increasingly difficult, an experiment of this kind does not, by modern
standards, require sophisticated detection equipment or calculations as it
is only measuring electron scattering. Again, further work is required.

\section{Conclusion} \label{sec:discuss} 
I have studied the long known but generally ignored result that
QCD with five or more colours has an altered vacuum state due to the limited 
dimensionality of space, a condition dubbed 'dimensional frustration'. It appears
to lead to a unified theory of the strong and electromagnetic interactions, which is
not the conventional approach to grand unification, but these two forces are the only
ones acting on the right-handed matter fields. Identification of 
the physical vacuum encounters an intractable set of non-analytic 
equations but further analysis was enabled by a well-motivated ansatz. 

Assuming the dual superconductor model, a range of confinement
scales emerged with one root vector (XXXO) being confined more strongly than all the rest,
while some others are less tightly confined (??OO). 
Gluons remaining at intermediate energy scales exhibit unconventional dynamics
because only some of them couple to the XXXO. An $SU(3)$ subset 
have stronger interactions among themselves at increasing length scales, suggesting
the separation of QCD dynamics at lower energy. In addition to a weakly confined
$U(1)$ and off-diagonal $SU(2)$ generators, there also emerged a single, unconfined
$U(1)$ gauge field consistent with the photon. The theory appears to be a unification
of QCD with electromagnetism. Such a unification, if consistent with experiment,
is of interest to the standard model in which right-handed matter fields only couple to those
two forces. 

Study of the matter field representations found that 
the fundamental representation $[1]$ comprises the three colours of down quark and two
more so-called invisible colours, named ultraviolet and infrared. These two colour charges
combined will neutralise each other, as do white combinations of the QCD colours. According
to the area law of the Wilson loop \cite{W74} 
combined with the non-Abelian Stokes theorem \cite{KT00a,KT00b}, the
asymmetric arrangement of the chromomonopole condensate gives the invisible colours, especially
ultraviolet, an extremely short confinement scale. 

The antisymmetric combination of $[1]$ with itself, denoted $[2]$, contains the anti-up
quark and the positron, as well as various combinations of the QCD and invisible colours.
(The remaining matter representations $[3],[4]$ are the complex conjugates of $[1],[2]$.)
Again the invisible-coloured quarks have a very short confinement scale and make little
contribution to the intermediate energy dynamics so that the effective theory reduces to
QCD and electromagnetism.

Much work remains to be done. The masses of gauge bosons coupling the QCD colours to
the invisible colours need to be calculated. The breaking off of QCD symmetry, both
through gluon interaction strength and quark confinement scales, suggests
a strong symmetry breaking that should render these bosons very massive.

Dimensional frustration is a natural, almost inevitable, means of generating a 
hierarchy in QCD with five or more colours without resorting to contrived symmetry breaking
methods such as the Higgs field. Even a simplistic analysis such as this finds a rich
phenomenology. 

The author thanks K.-I. Kondo for helpful discussions. This work was partially supported by a
fellowship from the Japan Society for the Promotion of Science (P05717), with hospitality
provided by the physics department of Chiba University.


\end{document}